\documentclass[epj,twocolumn]{webofc}
\usepackage[varg]{txfonts}   
\usepackage{textgreek}
\woctitle{MPGD2015}
\begin{document}
\title{Fast Timing for High-Rate Environments with Micromegas}
%
%

\author{Thomas Papaevangelou\inst{1}\fnsep\thanks{\email{thomas.papaevangelou@cea.fr}} \and
        Daniel Desforge\inst{1} \and
        Esther Ferrer-Ribas\inst{1} \and
        Ioannis Giomataris\inst{1} \and
        Cyprien Godinot\inst{1} \and
        Diego Gonzalez Diaz\inst{2} \and
        Thomas Gustavsson\inst{3} \and
        Mariam Kebbiri\inst{1} \and
				Eraldo Oliveri\inst{2} \and		
				Filippo Resnati\inst{2} \and 
				Leszek Ropelewski\inst{2} \and
        Georgios Tsiledakis\inst{1} \and
				Rob Veenhof~\inst{2} \and
        Sebastian White\inst{2,4} 
}

\institute{CEA, IRFU, Centre d'Etudes Nucléaires de Saclay, Gif-sur-Yvette 91191, France 
\and
           European Organization for Nuclear Research (CERN), Genève 1211, Switzerland 
\and
					 Laboratoire Interactions, Dynamiques et Lasers (LIDyL) CEA, CNRS, Université Paris-Saclay, 
Gif sur Yvette 91191, France
\and
           Princeton University, Princeton, New Jersey 08544, USA
          }

\abstract{The current state of the art in fast timing resolution for existing experiments is of the order of 100~ps on the time of arrival of both charged particles and electromagnetic showers. Current R\&D on charged particle timing is approaching the level of 10~ps but is not primarily directed at sustained performance at high rates and under high radiation (as would be needed for HL-LHC pileup mitigation). We demonstrate a Micromegas based solution to reach this level of performance. The Micromegas acts as a photomultiplier coupled to a Cerenkov-radiator front window, which produces sufficient UV photons to convert the $\sim$100~ps single-photoelectron jitter into a timing response of the order of 10-20~ps per incident charged particle. A prototype has been built in order to demonstrate this performance. The first laboratory tests with a pico-second laser have shown a time resolution of the order of 27~ps for $\sim$50 primary photoelectrons, using a bulk Micromegas readout. 
}
\maketitle
\section{Introduction}
\label{intro}
From 2025 onward, it is planned to operate the LHC with typically 140 collisions per proton bunch crossing (HL-LHC phase), which will greatly complicate the analysis, particularly in the forward regions, where it will be very difficult to link the tracks with the primary vertex (associated with the only interesting collision) and thus to prevent the formation of fake jets created by the random stacking or pile-up of tracks from several secondary vertices. A solution proposed to fight against this pile-up effect is to measure very accurately the time of arrival of the particles~\cite{refS}. A time resolution of a few tens of picoseconds with a spatial granularity better than 10~mm (depending on pseudo-rapidity) will be needed to obtain a fake jet rejection rate that is acceptable for physics analyses. Such performances should be obtained with a detector capable of withstanding the very important radiation levels expected in the HL-LHC era.

The ongoing R\&D on timing for charged particles is approaching a level of 10~ps \cite{RefJ} but is not primarily directed at sustained performance at high rates and under high radiation (as would be needed for HL-LHC pileup mitigation). Several technologies are presently under evaluation within the LHC collaborations: Silicon detectors with some built-in amplification, Silicon detectors without amplification and gaseous detectors. Gaseous detectors have the potential of very good timing performance, provided the amplification gap is small enough. They have also the advantages of being low-cost, simple to operate, and are potentially insensitive to irradiation effects. 

Modern photo-lithographic technology has enabled the development of a series of novel Micro Pattern Gaseous Detectors (MPGD): Micro-Strip Gas Chamber (MSGC), Gas Electron Multiplier (GEM), Micro-mesh Gaseous Structure (Micromegas), and many others revolutionizing cell size limits for many gas detector applications. The Micromegas detector fulfils the needs of high-luminosity particle physics experiments, and compared to classical gas counters, it offers intrinsic high rate capability, excellent spatial resolution and excellent time resolution ~\cite{refG1,refG2}.

\section{Micromegas for fast timing}
\label{sec-2}
\subsection{Detection principle}
\label{sec-21}

The concept of a Micromegas detector coupled to a front window that acts as Cerenkov-radiator, equipped with a photocathode, is examined in order to address its timing performance. The radiator should produce sufficient UV photons and should be equipped with an appropriate photocathode in order to create the necessary number of electrons to convert the expected $\sim$100~ps single-photoelectron time jitter into a timing response of 10-20~ps per incident particle. The photocathode has to be spark-resistant and must have good aging properties, appropriate for the high particle flux environment of HL-LHC.

\begin{figure}[h!]
\centering
\includegraphics[width=0.99\columnwidth]{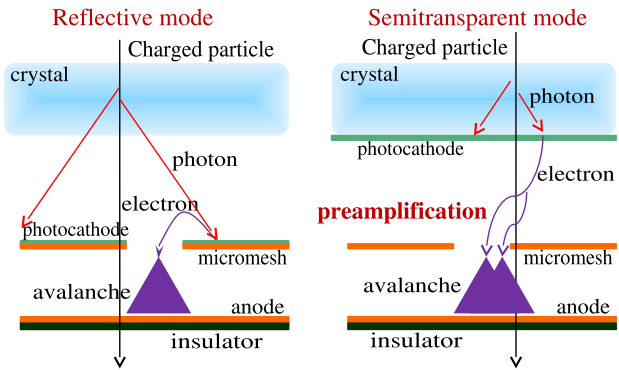}
\caption{Schematic view of the charged particle detection principle, in the two modes; Reflective mode: a thin CsI layer is deposited on top of the micromesh; Semitransparent mode: CsI is deposited on the bottom surface of the UV crystal, defining a drift gap with the micromesh.}
\label{fig-schem}       
\end{figure}

In gaseous detectors the time resolution is typically of the order of several ns because of the time jitter of primary ionization (for high energy MIPs) in the drift gap. Our strategy to improve the time response is to minimize this jitter by creating simultaneously all primary electrons at about the same distance from the amplification region and by minimizing the drift space. The primary electrons are created by the Cerenkov light which is emitted by the charged particle when it crosses a 2-3~mm MgF$_{2}$ crystal. A thin CsI photocathode is deposited on the micromesh (reflective mode) or on the bottom of the crystal (semitransparent mode) as shown in figure \ref{fig-schem}, and it is used to convert UV photons to electrons. The time spread of the Cerenkov light ($\sim$10~ps) is too small to limit the projected time accuracy.

In the semitransparent mode the photocathode is deposited at the bottom face of the MgF$_2$ crystal at a distance of 100-300~\textmu m from the micromesh to limit longitudinal diffusion. A strong field can be applied in the drift gap allowing some pre-amplification. The advantages of this configuration are the high single-electron sensitivity due to higher total gain, the possibility to use woven mesh Micromegas (bulk technology \cite{refBu}) and the fabrication simplicity. However, the partial exposure of the photocathode to the ion backflow may cause aging and robustness issues.
In the reflective mode the photocathode is deposited on the Micromegas mesh and no strong drift field is needed. The main advantages of this method is the very effective protection of the photocathode from the ion backflow (better aging properties, robustness) and the complete suppression of gas ionization in the drift region. However, this mode is limited to flat mesh detectors (Microbulk technology is the more appropriate \cite{refMB}), while the internal reflection of the Cerenkov light on the crystal-gas interface might limit the efficiency.

\subsection{Photoelectron production}
\label{sec-22}
A MIP passing through a crystal will emit:
	 \[
 \frac{d^2N}{dEdx} = \frac{\alpha^2z^2}{r_e m_e c^2}\sin^2 \theta_c \approx \frac{370}{eV cm}\left( 1 - \frac{1}{n^2(E)}\right)
\]
photons. If $L$ is the thickness in cm, $T(E)$ is the transmission of the crystal and $QE(E)$ the quantum efficiency of the photocathode, the number of produced photoelectrons will be:
	\[
	N_{p.e.} \approx 370 L \int { T(E) QE(E) \left(  1 - \frac{1}{n^2(E)} \right) dE}
\]
  
\begin{figure}[h!]
\centering
\includegraphics[width=0.95\columnwidth,clip]{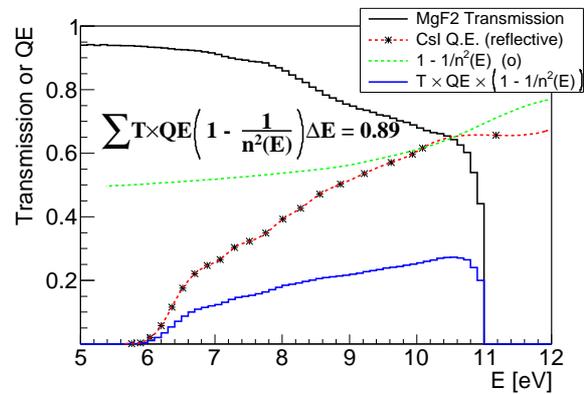}
\caption{Transmission, refractive index of MgF$_2$ and quantum efficiency of CsI in reflective mode which were used for the calculation of the number of photoelectrons.}
\label{fig-crystal}       
\end{figure}

For a MgF$_2$ crystal and a CsI photocathode with $QE(E)$ taken from \cite{refQE}, $T(E)$ from \cite{refT} and $n(E)$ from \cite{refn}, we expect $370\times0.89=330$~p.e./cm (figure \ref{fig-crystal}). However, this $QE$ concerns operation in reflective mode. In the semitransparent mode it is expected to be significantly lower (factor 2-3), so, for a 3~mm MgF$_2$ crystal we would expect 30-50 photoelectrons.  

\subsection{Diffusion in the drift region}
\label{sec-23}
Since the time jitter in photoelectron creation from the Cerenkov light is small (<10~ps), the dominant effect limiting the time resolution will be the diffusion of the photoelectrons in the drift region. 

\begin{figure}[h!]
\centering
\includegraphics[width=0.95\columnwidth,clip]{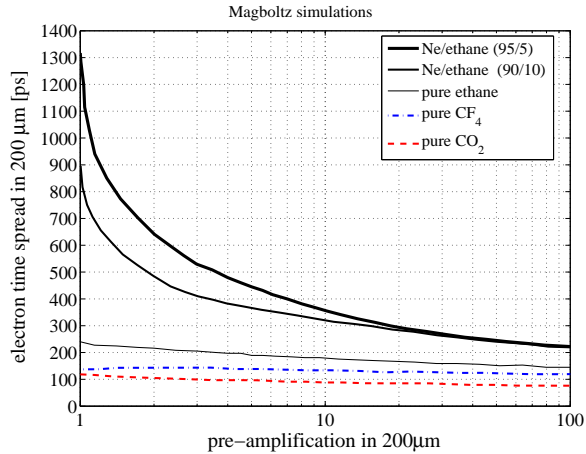}
\caption{Electron time spread due to diffusion for several gases as a function of the achieved pre-amplification in a 200 \textmu m gap.}
\label{fig-sim}       
\end{figure}

In the case of the reflective photocathode, the jitter in the drift path will reflect the difference in the path depending on the creation point on the mesh and its distance to the amplification gap, so fine pitch meshes are preferable.
Simulations were carried out to investigate the semitransparent mode. Magboltz simulation results are shown in figure \ref{fig-sim} where the time spread due to electron diffusion in a 200~\textmu m drift gap is plotted as function of the preamplification for several gas mixtures. A first assessment based on Magboltz thus indicates that a single photoelectron time resolution of 100-200~ps could be reachable for relatively small (or even in the absence of) pre-amplification. 

Compared to noble-gas admixtures, pure quenchers show potentially a smaller time spread, but demand stronger electric fields in order to achieve the same gain. The optimum selection will be a tradeoff between the minimum time jitter and the maximum signal to noise ratio.

\subsection{The Micromegas prototype}
\label{sec-23}

\begin{figure}[h!]
\centering
\includegraphics[width=0.99\columnwidth]{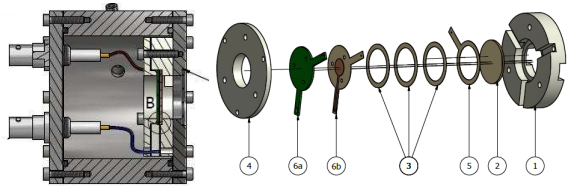}
\caption{Schematic view of the detector chamber and expanded view of the Micromegas - spacers - crystal assembly.}
\label{fig-layout}       
\end{figure}

A prototype detector has been constructed (figure ~\ref{fig-layout}) in order to study the timing properties of the proposed scheme. It has been designed to be compatible with operation in both semitransparent and reflective mode. 

The Micromegas was constructed using both Bulk and Microbulk technologies~\cite{refBu,refMB}. It was a single-anode detector of 1~cm diameter (bigger than the maximum pixel size of a possible tracker). The amplification gap was 50~\textmu m for the Microbulk and 128~\textmu m for the bulk. Kapton rings, 50~\textmu m thick, were used in order to support the MgF$_2$ or Quartz crystal on which the photocathode is deposited and define the drift gap of 200~\textmu m. The top ring was metalized on the top surface, in order to apply the bias voltage to the photocathode. A thin (10~nm) Al layer was deposited on a quartz crystal and served as photocathode. This simplified configuration is sufficient to study the timing properties of the detector, using a very fast UV laser. For charged particle detection, the quartz has to be replaced with MgF$_2$, while the photocathode will be a CsI layer (100-300~nm) deposited on a metallic substrate (5-50~nm). The exact configuration will be optimized in terms of aging properties and of the maximization of the number of produced photoelectrons. 

The Micromegas-crystal assembly is encased in a stainless-steel chamber. A quartz window, transparent to the UV laser, is mounted on the top surface of the chamber. The chamber and the components were properly chosen and special precaution was taken while mounting the detector in order to be appropriate for operating without gas circulation, to facilitate thus the measurements at laser and particle beam facilities. 

\subsection{Performance of the prototype}
\label{sec-24}
The gas mixture was chosen to be 90\% Ne plus 10\% Ethane, a mixture that gives high gain with good energy resolution and has shown good behavior for sealed mode operation \cite{refFF}, though it is not optimal for timing. We used a pulsed deuterium lamp in order to monitor the gain stability without gas circulation. Data were taken over one week showing only random fluctuations (arising from the lamp intensity) and no significant drop of the gain, demonstrating the possibility to perform measurements without renewing the gas. 

\begin{figure}[h!]
\centering
\includegraphics[width=0.925\columnwidth]{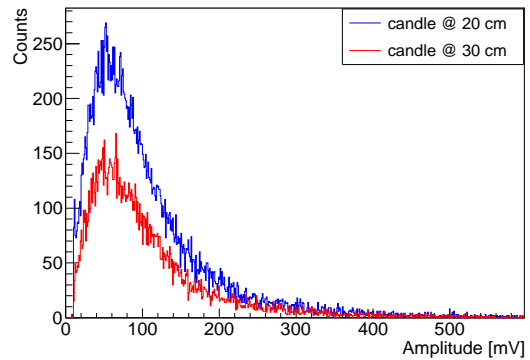}
\caption{Pulse amplitude spectra for a candle placed at $\sim$20~cm and $\sim$30~cm distance from the detector with a charge preamplifier. The shape of the spectrum does not change, a clear indication for single electron pulses.}
\label{fig-candle}       
\end{figure}

The sensitivity of the detector to single photons was verified using a candle flame, which is emitting continuously light from the IR down to the VUV, the spectrum stopping at $\sim$200~nm due to the absorption of the air \cite{refFF}. The quantum efficiency of the aluminium is sufficient to have several photoelectrons per second when the distance to the flame is not very long. Two spectra taken at a distance of 20 and 30 cm for the same time duration are shown in figure~\ref{fig-candle}. The amplitude distribution does not change with the distance, while the rate does, proving that the pulses correspond to single photons. 
The resulting very good energy resolution is due to the single stage amplification in the Micromegas gap.

\section{Time resolution measurements}
\label{sec-3}

The time response of the bulk Micromegas detector in semitransparent mode was studied at LIDyL laboratory (CEA/Saclay). An Optical Parametric Oscillator (Coherent MIRA-OPO), pumped by a Ti:sapphire laser (Coherent MIRA 900) provided 120~fs pulses at 550~nm. The repetition rate was varied between 9 kHz and 4.75 MHz by using a Pulse-Picker (Coherent 9200). The second harmonic at 275 nm was generated in a 5 mm BBO crystal. The typical output energy was 0.15-1~pJ for a beam diameter < 1~mm.

Though the extraction efficiency for the Al photocathode at those wavelengths is small, the intensity of the laser (up to $10^6$ photons/pulse) allows for the production of a sufficient number of photoelectrons (up to 1000 p.e./pulse for 10 nm Al, depending also on the beam tuning and the applied Micromegas drift field). The number of photoelectrons per laser pulse was controlled by using a series of light attenuators: fine electroformed nickel meshes (100-2000 LPI) with optical transmission varying between 10\% - 25\%, giving attenuation factors of 4, 5, 10 and their combinations. The pitch of the meshes was varied in order to avoid systematic effects when combining several of them.  

\begin{figure}[h!]
\centering
\includegraphics[width=0.95\columnwidth,clip]{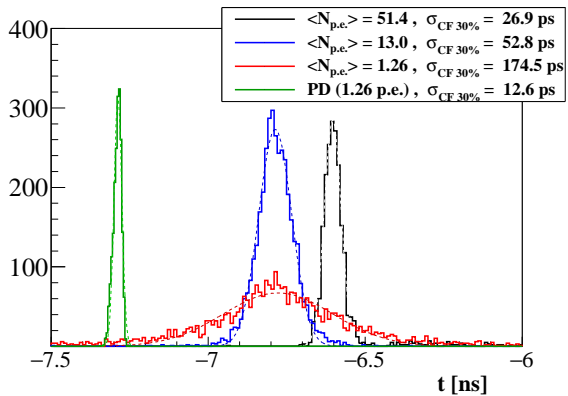}
\caption{Distribution of the time at the 30\% fraction of the amplitude for the Micromegas pulses, for different mean number of photoelectrons. The time spread of the photodiode is also shown.}
\label{fig-analysis}       
\end{figure}

The laser beam was split in two, part of it illuminating our prototype and the other part an ultra fast Si detector (Thorlabs DET10A/M). Both signals were recorded by a 2 GHz oscilloscope operating at 10~GS/s. The time jitter of the photodiode was of the order of 13~ps, being appropriate as trigger and as reference for the measurement. The signal from the Micromegas was filtered for high frequency noise using a moving average. The photocathode and the filtered Micromegas signal times were calculated by taking a constant fraction of each signal. Best results were obtained for a 30\% fraction of the amplitude (figure~\ref{fig-analysis}).    

\begin{figure}[h!]
\centering
\includegraphics[width=0.99\columnwidth]{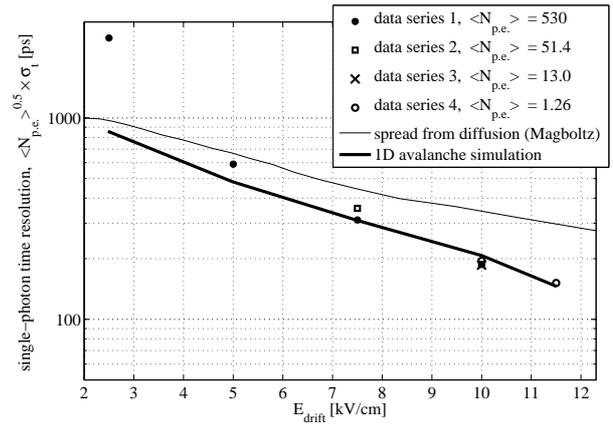}
\caption{Dependence of the measured time resolution with the drift field, scaled to the single photoelectron case. The difference between the thin and thick lines indicates the improvement due to pre-amplification, according to a stochastic 1D avalanche model.}
\label{fig-drift}       
\end{figure}

\begin{figure}[h!]
\centering
\includegraphics[width=0.99\columnwidth]{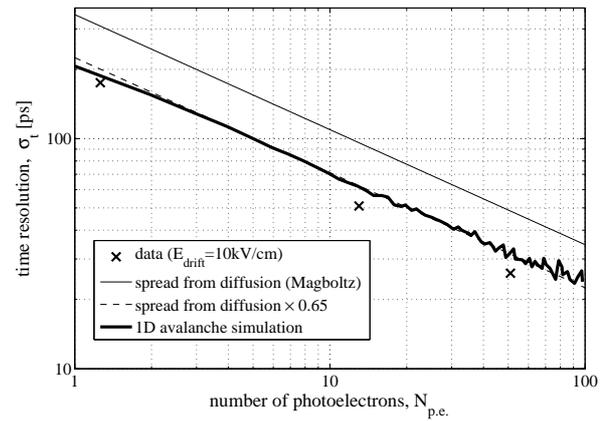}
\caption{Dependence of the measured time resolution with the mean number of photoelectrons, for fixed amplification and drift fields. A resolution of 200 ps per single photoelectron and 27~ps for ~50 photoelectrons has been achieved. }
\label{fig-npe}       
\end{figure}

Data were taken for different values of the drift field in combination with the attenuators, in order to maintain the signal within a reasonable dynamic range. The time resolution improves with the drift field intensity due to the decrease of the longitudinal diffusion but also due to the preamplification at high field values. The results are summarized in figures~\ref{fig-drift} and~\ref{fig-npe}, in comparison with a stochastic 1D avalanche simulation (that will be detailed in a forthcoming publication). 

\section{Conclusions}
\label{sec-4}
A Micromegas photodetector, equipped with a UV-transparent crystal and a metallic photocathode deposited on the bottom of the crystal has been constructed in order to investigate the potential of MPGDs for fast timing tracking of minimum ionizing particles. The aim of the first tests was the investigation of the intrinsic time characteristics of the detector. A very fast UV laser (time spread 120~fs) was used to produce the photoelectrons in a controlled way. With a gas mixture consisting of 90\% Ne and 10\% Ethane we measured a time resolution of the order of 200~ps for single photoelectrons and 27~ps for 50 photoelectrons. Simulations show that this already satisfactory result can be further improved by optimizing the gas mixture. A more exhaustive description of these results will be published elsewhere. Tests on a particle beam, with a MgF$_2$ crystal and CsI photocathode will be made in 2016. 

The current project is carried out within the framework of the RD51 common Project "Fast Timing for High Rate Environments: A Micromegas Solution".


\begin{thebibliography}{}
\bibitem{refS}
S. White, proc Picosecond Workshop, Clermond-Ferrand 2014, arXiv:1409.1165 [physics.ins-det].

\bibitem{RefJ}
C. Williams et al., Nucl.Instrum.Meth. \textbf{A594} (2008) 39-43. 

\bibitem{refG1}
Y. Giomataris, P. Rebourgeard, J.P. Robert, G. Charpak, Nucl.Instrum.Meth. \textbf{A376} (1996) 29. 

\bibitem{refG2}
I. Giomataris,  Nucl.Instrum.Meth. \textbf{A419} (1998) 239.

\bibitem{refBu}
 I. Giomataris et al., Nucl.Instrum.Meth. \textbf{A560} (2006) 405-408 

\bibitem{refMB}
S. Andriamonje et al., JINST \textbf{5} (2010) P02001

\bibitem{refQE}
J. Séguinot et al., Nucl.Instrum.Meth. \textbf{A297} Issues 1-2 (1990) 133-147.

\bibitem{refT}
http://www.korth.de/index.php/162/items/21.html

\bibitem{refn}

P. Laporte et al., Optical Society of America Journal (ISSN 0030-3941), \textbf{73}, (1983) 1062-1069.

\bibitem{refFF}
A. Peyaud et al., Nucl.Instrum.Meth. \textbf{A787} (2015) 102-104. 
\end{thebibliography}
\end{document}